 \documentclass[12pt,preprint]{aastex}

 \usepackage{emulateapj5}

\slugcomment{to appear in the Astrophysical Journal, Letters}

\shorttitle{Light Curve of V1494 Aql}
\shortauthors{Hachisu et al.}

\begin{document}

\title{Detection of two-armed spiral shocks on the accretion disk 
of the eclipsing fast nova V1494 Aquilae}


\author{Izumi Hachisu}
\affil{Department of Earth Science and Astronomy, 
College of Arts and Sciences, University of Tokyo,
Komaba, Meguro-ku, Tokyo 153-8902, Japan} 
\email{hachisu@chianti.c.u-tokyo.ac.jp}

\author{Mariko Kato}
\affil{Department of Astronomy, Keio University, 
Hiyoshi, Kouhoku-ku, Yokohama 223-8521, Japan} 
\email{mariko@educ.cc.keio.ac.jp}

\and

\author{Taichi Kato}
\affil{Department of Astronomy, Kyoto University, 
Sakyo-ku, Kyoto 606-8502, Japan} 
\email{tkato@kusastro.kyoto-u.ac.jp}




\begin{abstract}
     We have modeled the unusual orbital light curve of V1494 Aquilae 
(Nova Aquilae 1999 No.2) and found that such an unusual
orbital light curve can be reproduced when there exist two-armed, 
spiral shocks on the accretion disk.  V1494 Aql is a fast classical
nova and found to be an eclipsing system with the orbital period of 
0.1346138 days in the late phase of the nova outburst.  Its orbital 
light curve shows a small bump at orbital phase 0.2, a small dip
at 0.3, sometimes a small bump at 0.4, and a large bump at $0.6-0.7$
outside eclipse.  Such a double- or triple-wave pattern outside eclipse
has never been observed even though overall patterns look like some
supersoft X-ray sources or eclipsing polars.
We have calculated orbital light curves including the irradiation effects
of the accretion disk and the companion by the hot white dwarf.
These unusual patterns can be reproduced when we assume two-armed
spiral shocks on the accretion disk.  Especially, triple-wave patterns
are naturally obtained.  This result strongly suggests the existence of
two-armed spiral shocks on the accretion disk 
in the late phase of the nova outburst.
\end{abstract}




\section{INTRODUCTION}
     Angular momentum transport plays an essential role 
in accretion disks of cataclysmic variables.
Two mechanisms have been proposed so far: 
one is the turbulent viscosity as adopted in the
standard accretion disk model of \citet{sha73} and the other is 
the direct dissipation by tidal spiral shocks on the accretion disk
as first demonstrated by \citet*{saw86}.  The turbulent viscosity 
is a local physical process while the tidal spiral shocks
have a global structure on the accretion disk. 
Therefore, we have a chance to observe a global spiral 
shock structure when they play an essential role in the angular 
momentum transport of the accretion disk.  
Such an observational evidence first came
from the Doppler maps of the dwarf nova IP Peg outburst
\citep*[e.g.,][]{ste97}.  We have long believed that tidal spiral
structures can be detected even in orbital light curves of 
cataclysmic variables if the spiral patterns are prominent.
At last we find such an evidence of spiral patterns on 
the accretion disk from the orbital light curves of V1494~Aquilae.

     V1494~Aquilae (Nova Aquilae 1999 No.2) was discovered by 
A. Pereira on 1999 December 1.785 UT at $m_V \sim 6.0$
\citep{per99}.  It reached the visual maximum brightness of
$m_V \sim 4.0$ on Dec 3.4 UT ($t_0=$JD 2,451,515.9$\pm 0.1$)
and decayed at the rate of $t_2= 6.6 \pm 0.5$ days and 
$t_3= 16 \pm 0.5$ days \citep{kis00}.  Early phase spectra were taken
by \citet{fuj99} and \citet{aya99}, which show P-Cyg profiles
of hydrogen Balmer lines with a blueshifted component of
1020 km~s$^{-1}$ and 1200 km~s$^{-1}$, respectively.
Thus, V1494~Aql has been classified as a fast nova.
\citet{kat04t} have summarized overall development of the V1494~Aql
light curve until autumn of 2003 (for about 3 years).  Detailed
spectral features have been reported by \citet{iij03}.

A short-term periodic modulation was first
reported by \citet{nov00}, who found 0.03 mag variations
with a period of 0.0627 days from their 2000 June observation.
\citet{ret00} analyzed 31 nights of CCD photometry
during 2000 June-August and suggested a periodicity of
0.13467(2) days.  The full amplitude of the variation in the $R$-band
increased from 0.03 mag in June to 0.07 mag in August.  Their folded
light curve shows a double-wave structure with a shallower dip at phase
0.5, with about half the amplitude of the main periodicity.
\citet{bos01} reported, based on unfiltered and $R$-band CCD photometry
obtained on 12 nights during 2001 June-July, a robust
change in the shape and amplitude of the 0.13467 days period.
It had an eclipse shape with depth about 0.5 mag.
A second shallow eclipse (about 0.1 mag deep) at phase 0.5 can be seen.
\citet{bar03} refined, based on $V$-band photometry during June and
September of 2002, the orbital period of 0.1346141(5) days.
They pointed out that the orbital light curve is quite unusual
in the sense that it is not similar to ordinary cataclysmic
variables with a hot spot on the accretion disk.
\citet*{pav03} made a multicolor photometry and concluded that
the eclipse depth is deeper in longer wave lengths (deeper in
$I$-band than in $V$-band).  \citet{pav03} further discussed
that such an eclipsing characteristic (deeper in longer wave lengths)
cannot be explained by the accretion disk or the irradiation
effect of the companion because these two light sources have
the opposite nature (deeper in short wave lengths).
Thus, they suggested, as a model of V1494~Aql orbital light curve,
a self eclipsing accretion column in a magnetic polar. 

\citet{kat04t} provided a time-development of the orbital light curves
among 2001 November-December, 2002 August, and 2003 June.  
It sometimes shows a triple-wave structure as well as a double-wave
structure.  Kato et al. suggested that some structures (probably
on the accretion disk) fixed in the binary rotational frame are
responsible for the orbital light curve because of the stability
of out-of-eclipse light curve patterns.  Very recently, 
\citet*{kis04} pointed out that a close companion to V1494~Aql
(located $1\farcs4$ southwest)
is brighter than V1494~Aql itself in the very late phase of the   
outburst.  They corrected the brightness of the orbital light
curve and found that the depth of the eclipse is about twice
deeper in magnitudes than before correction.  

     The depth of the primary eclipse of V1494~Aql 
has become deeper as the nova 
decayed.  Such a feature was also observed in the 
recurrent nova CI~Aql 2000 outburst \citep[e.g.,][]{mat01}, in
which the irradiation of the accretion disk plays an essential role 
\citep{hac03ka}.  Therefore, we expect the accretion disk 
in V1494~Aql is also responsible for 
the unusual wavy structure of the orbital light curve.
In this Letter, we model the orbital light curve.
Almost the same orbital light curve model as 
in the supersoft X-ray sources is adopted
\citep*[e.g.,][]{sch97, hac03kb, hac03kc}, which includes 
the irradiation effects of the accretion disk and of the companion.
In addition, we assume two-armed spiral structures on the accretion 
disk to reproduce a triple-wave structure outside eclipse.
In \S 2, our numerical model for V1494~Aql
is briefly introduced and summarized.  The numerical
results are given in \S 3.  Discussion follows in \S4.

\placefigure{v1494_figure_late_i785}
\placetable{system_parameters_V1494_Aql}

\section{THE LIGHT CURVE MODEL}
     Our binary model is illustrated 
in Figure \ref{v1494_figure_late_i785},
which consists of a $0.3~M_\sun$ main-sequence star (MS) filling
its Roche lobe, a $1.0~M_\sun$ white dwarf (WD), 
and a disk around the WD.
The mass of the WD is roughly estimated from the decline rate
of the decay phase \citep[e.g.,][]{hkkm00, hac01ka, hac01kb},
details of which will be published elsewhere. 
The mass of the companion is estimated from
the mass-period relation for cataclysmic variables 
\citep[e.g.,][]{pat84, war95}.
A circular orbit is assumed.
Its ephemeris has recently been refined by \citet{kat04t}.  
We use this ephemeris, i.e.,
\begin{equation}
t(\mbox{BJD})= 2,452,458.3230 + 0.1346138 \times E,
\label{new_ephemeris}
\end{equation}
at eclipse minima. 

We also assume that the surfaces of the WD,
the MS companion, and the accretion disk 
emit photons as a blackbody at a local temperature 
which varies with position.
For the basic structure of the accretion disk,
we assume an axi-symmetric structure with the size and thickness of
\begin{equation}
R_{\rm disk} = \alpha R_1^*,
\label{accretion-disk-size}
\end{equation}
and
\begin{equation}
h = \beta R_{\rm disk} \left({{\varpi} 
\over {R_{\rm disk}}} \right)^\nu,
\label{flaring-up-disk}
\end{equation}
where $R_{\rm disk}$ is the outer edge of the accretion disk,
$R_1^*$ is the effective radius of the inner critical Roche lobe 
for the WD component,
$h$ is the height of the surface from the equatorial plane, and
$\varpi$ is the distance on the equatorial plane 
from the center of the WD.  Here, we adopt $\nu=2$.
We obtain our modeled two-armed spiral structures by multiplying
$h$ with $z_{\rm height}$, as defined below:
\begin{equation}
z_1 = \max \left(1, {{\xi_1} \over {\sqrt{(\varpi/R_{\rm disk} 
-\exp(-\eta (\phi - \delta)))^2 
+ \epsilon^2}}} \right),
\label{spiral-shock-pattern-1}
\end{equation}
\begin{equation}
z_2 = \max \left(1, {{\xi_2} \over {\sqrt{(\varpi/R_{\rm disk} 
-\exp(-\eta (\phi - \delta - \pi)))^2 
+ \epsilon^2}}} \right),
\label{spiral-shock-pattern-2}
\end{equation}
\begin{equation}
z_{\rm height} = \max(z_1, z_2).
\label{spiral-shock-pattern-total}
\end{equation}
We further have a slope at the disk edge defined by
$z-10(z-0.25)(\varpi /~R_{\rm disk}-0.9)$ for $0.9
< \varpi/R_{\rm disk} < 1.0$.  The above various disk parameters
are assumed to be $\epsilon = 0.1$, $\xi_1 = \xi_2 = 0.40$, 
$\eta = 0.41$, $\delta = 65\arcdeg$, $\alpha= 0.8$, $\beta = 0.095$
for the disk shape of Figure \ref{v1494_figure_late_i785}.  
Here, $\xi_1$ and $\xi_2$ specify the amplitudes of
two spirals so that $\xi_1 = \xi_2 ~(= 0.40)$ means just an anti-symmetric
structure of spirals with the same height, 
$\eta$ determines the inverse of the pitch angle of logarithmic
spirals, $\delta$ is the position angle of the spirals 
against the binary components, $\epsilon$ denotes the width of 
the spiral pattern and represents the height of the spiral 
against the thickness of the disk together with $\xi_1$ and $\xi_2$, 
i.e., $z_{\rm height} = \max(\xi_1,~\xi_2)/\epsilon \sim 4$ 
at the edge of the disk for the disk shape of 
Figure \ref{v1494_figure_late_i785}.
In our light curve model, we mainly change four parameters,
i.e., the disk thickness ($\beta$), the inverse of the pitch angle
of spirals ($\eta$), the position angle ($\delta$), 
and the heights of two spirals ($\xi_1 = \xi_2$).
The other parameters that specify the disk shape
are all fixed to be the above values.
These parameters are roughly determined to mimic the results of
3D simulation model of accretion disks \citep*[e.g.][]{mak00}. 

The luminosity of the WD is assumed to be $750-3,000 ~L_\sun$
because we do not know the exact WD luminosity at the time of
observation.  The disk and the companion star are strongly 
irradiated by the hot WD.  The surfaces of the WD, the disk, 
and the companion star are divided into many patches as shown 
in Figure \ref{v1494_figure_late_i785}.  Here we assume that 
each patch emits photons as a blackbody (with a single temperature).
Each patch of the disk or of the companion is irradiated 
by visible (front side) patches of the hot WD.  
The total luminosity of the irradiated disk and companion is 
calculated by summing up the contributions from all patches.  
The irradiation efficiency is the same as that of \citet{sch97},
i.e., 50\% of the irradiated energy is emitted by photon
but the residual 50\% is converted into thermal energy of gas.
The accretion luminosity of the disk is also numerically
included, although its contribution to the optical light is much
smaller than that of the irradiation effect
\citep*[see discussion of][]{hac01ka, hac01kb, hac03a}.
The numerical method adopted here
was fully described in \citet{hac01kb, hac03ka, hac03kb, hac03kc}.
The inclination angle ($i$) of the binary is a free
parameter that is determined from our light curve fitting. 
The original temperature of the companion star is assumed 
to be 3,000 K, but the shape of the orbital light curve
is hardly changed even if we take 2,000 K or 4,000 K for
the original temperature of the MS companion.  The adopted
system parameters are summarized in Table 
\ref{system_parameters_V1494_Aql}.

\placefigure{vmag_wd1000m03_quiescence_kiss}

\section{NUMERICAL RESULTS}
       The best fit model is plotted 
in Figure \ref{vmag_wd1000m03_quiescence_kiss} 
by a thick solid line.  The orbital modulations of V1494~Aql have
been reported by several groups.  Here, we adopt data of two groups 
\citep{kat04t, kis04} to fit with our modeled light curves.
Kato et al.'s (square) data have been corrected
by using Kiss et al.'s (circle) data,
because Kato et al.'s original data include the light 
from the close companion.

We have changed five parameters 
($\beta$, $\eta$, $\delta$, $\xi_1 = \xi_2$, $i$)
independently and calculated the total of $\sim 1300$ orbital light 
curves.  It takes about 6 hours to calculate one orbital light curve 
on a $2.4$ GHz Pentium 4 processor, because the surface patch elements 
are $64 \times 128 ~(\theta \times \phi)$ for the companion surface,
$64 \times 128 \times 2 ~(\varpi \times \phi \times$ (up and down sides))
for the disk, and $16 \times 32 ~(\theta \times \phi)$ for 
the WD, and the total of 129 steps for one orbital period.  
We have used 5 CPUs, so that it took the total computation time 
of about 1600 hours, i.e., $(6 \times 1300)/$5 CPUs.
We have found, by the least square method, the best fit model
for out-of-eclipse (orbital phase $0.1-0.9$),
i.e., $\beta=0.095$, $\eta=0.41$, $\delta=65\arcdeg$,
$\xi_1 = \xi_2 = 0.40$, $i=78.5\arcdeg$.
For the best fit model, we increase the number of patches
for the disk up to $128 \times 256 \times 2 
~(\varpi \times \phi \times$ (up and down sides)) in order to
calculate a rather smooth light curve.  Such a high quality 
light curve is shown in Figure \ref{vmag_wd1000m03_quiescence_kiss}.

     The triple-wave structure of out-of-eclipse orbital light curve
is reasonably and naturally reproduced 
with our spiral shock pattern model.  Both the thickness 
of the accretion disk and the pitch angle of spirals adopted here
are roughly consistent with the 3D simulations \citep[e.g.,][]{mak00}. 
The present results strongly support the two-armed spiral pattern
on the accretion disk.  We hope a spectroscopic confirmation of 
this spiral pattern by Doppler maps for V1494~Aql.

\placefigure{vmag_wd1000m03_quiescence_kiss_2}

\section{DISCUSSION}
\begin{figure}
\plotone{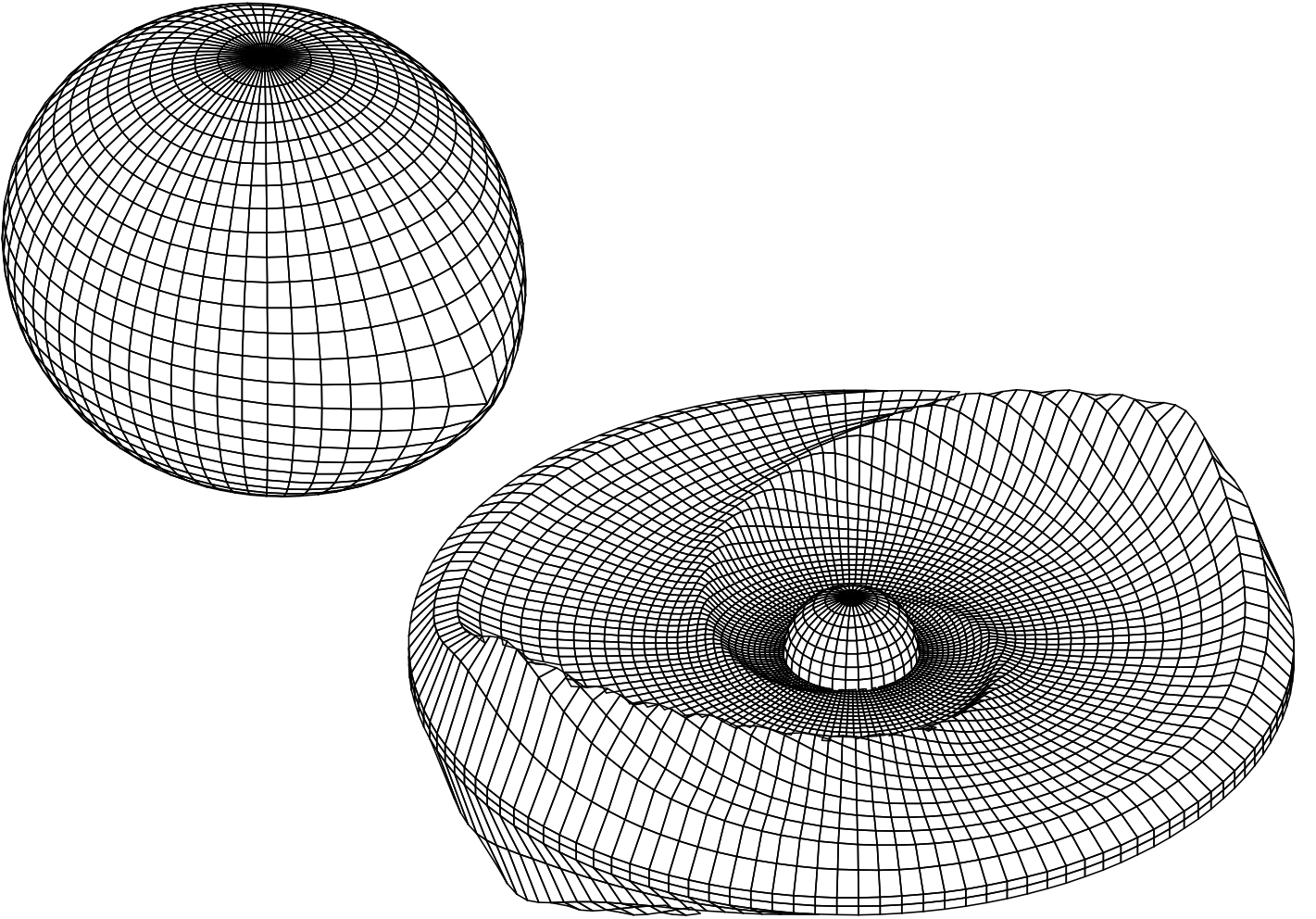}
\caption{
Our V1494~Aql model is illustrated.
The cool component ({\it far and left side}) is an 
MS companion ($0.3~M_\sun$) filling up its inner critical Roche lobe.  
The north and south polar areas of the cool component are 
irradiated by the hot component ($1.0~M_\sun$ WD, 
{\it near and right side}).  The separation is $a= 1.21 ~R_\odot$; 
the effective radii of the inner critical Roche lobes are
$R_1^*= 0.59 ~R_\odot$, and $R_2^*= R_2= 0.34 ~R_\odot$, 
for the primary WD and the secondary MS companion, respectively.
A two-armed spiral pattern is shown on the accretion disk.
The WD surface is artificially enlarged to easily see it.
\label{v1494_figure_late_i785}}
\end{figure}

\begin{figure}
\plotone{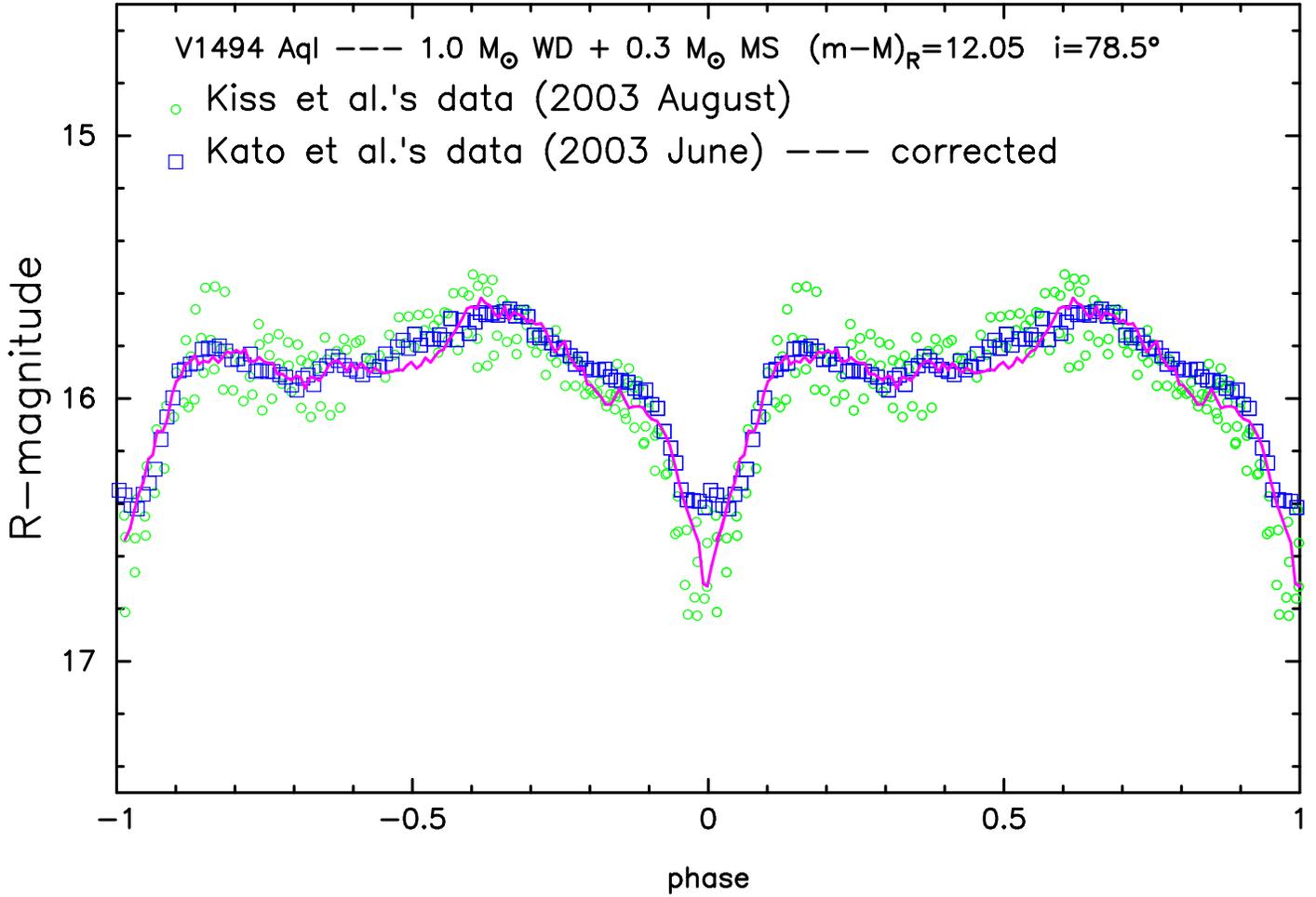}
\caption{
Calculated $R_c$ light curves are plotted against the binary phase 
(binary phase is repeated twice from $-1.0$ to $1.0$) together with
the observational points of \citet[][square]{kat04t} and 
of \citet[][circle]{kis04}.  A thick solid line denotes
$R_c$ light curve for the best fit model.
The apparent distance modulus is $(m-M)_R= 12.05$ for 
the WD luminosity of $L_{\rm WD}= 3,000~L_\sun$.
The adopted system parameters are listed in Table
\ref{system_parameters_V1494_Aql}.
\label{vmag_wd1000m03_quiescence_kiss}}
\end{figure}

\begin{figure}
\plotone{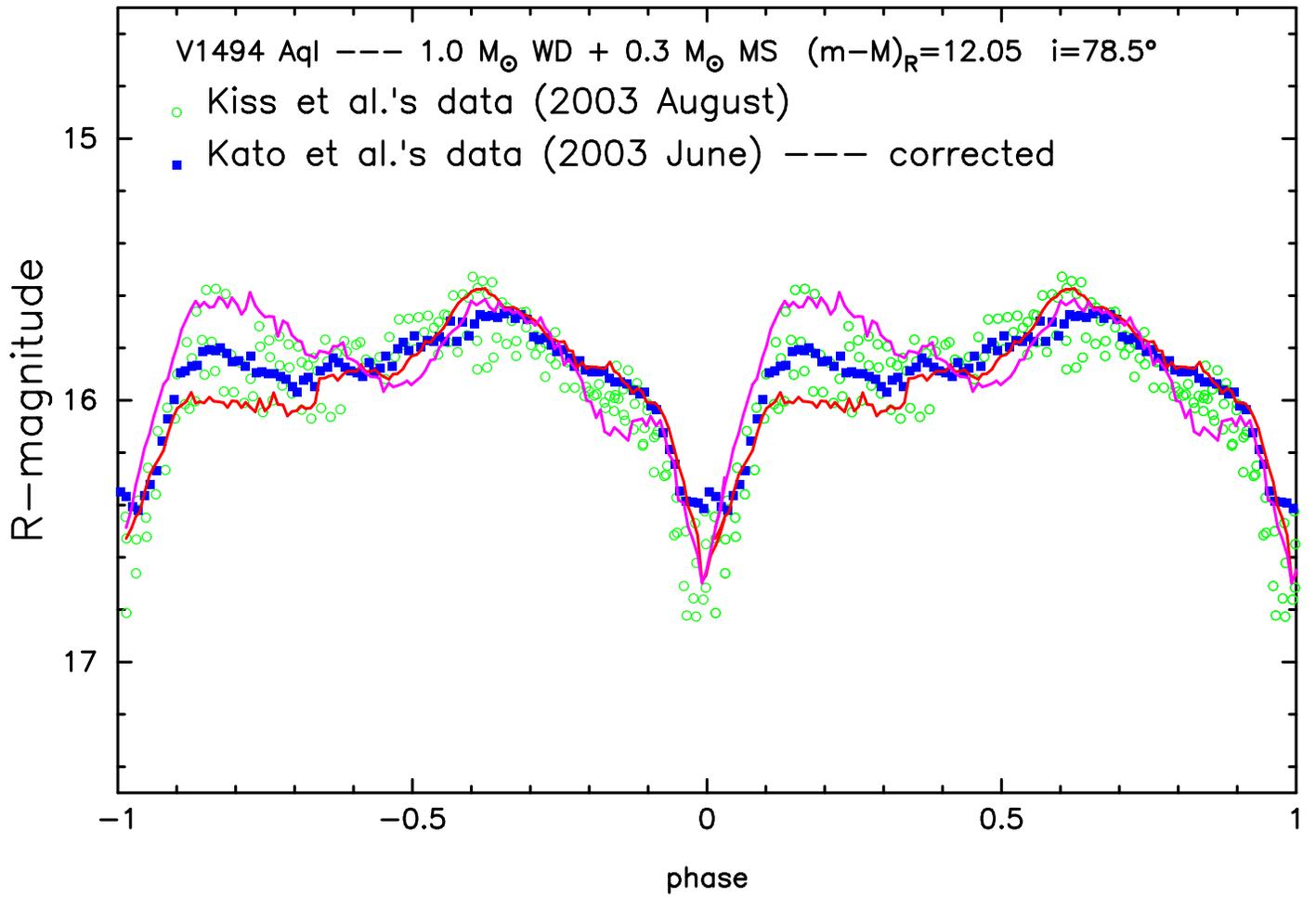}
\caption{
Same as in Fig. \ref{vmag_wd1000m03_quiescence_kiss} but for
variations of numerical models with different heights of spirals.
The variations appear at orbital phase of $0.1-0.3$
but the peak at phase $\sim 0.6-0.7$ is robust.  See text for details.
\label{vmag_wd1000m03_quiescence_kiss_2}}
\end{figure}

The orbital light curve varies from night to night as shown in
Figure 3 of \citet{kat04t}.  Such a scatter is also clearly shown 
in the present Figures 2 and 3 for Kiss et al.'s (2004) data.  
To reproduce this variation, we have changed parameter $\xi_1$
and $\xi_2$ independently 
(see Fig. \ref{vmag_wd1000m03_quiescence_kiss_2}). 
The variation appears in orbital phase of $0.1-0.3$
and the upper and lower bounds for these variations can be well
reproduced by the models of $\xi_1=0.26$, $\xi_2=0.44$ 
for the upper bound,
and $\xi_1=0.44$, $\xi_2=0.24$ 
for the lower bound.  Here, $\xi_1$ and $\xi_2$ represent 
the height of the spiral arm in the rear and front side 
on the disk of Figure \ref{v1494_figure_late_i785}, respectively.
The other parameters are the same as those in
Figures \ref{v1494_figure_late_i785} 
and \ref{vmag_wd1000m03_quiescence_kiss}.

It is remarkable that the peak at orbital phase 
$\sim 0.65$ is rather stable and never disappears. 
This is consistent with the robustness of the period
for the out-of-eclipse shape as emphasized by \citet{kat04t}.

     \citet{pav03} discussed the main light source for the wavy
structures of the out-of-eclipse orbital light curve.
They argued, as a light source, (1) accretion disk, 
(2) ellipsoidal shape of the companion, (3) reflection
(irradiation) effect of the companion by the hot white dwarf,
(4) intermediate polar activity, and (5) polar activity self-eclipsed
by the accretion column.  They rejected possibilities
(1)-(4) and suggested a viable model of polar activity.
Their reasons rejecting possibilities (2) and (4) are reasonable but
not for possibilities (1) and (3) as discussed below. 

\citet{pav03} rejected the possibility of accretion disk by
pointing out two reasons: 
(i) the maximum duration of the eclipse of the disk
limited by the Roche lobe size and it cannot exceed a quarter of
the orbital phase while the overall duration of the eclipse is
as large as 0.45 of the orbital period.  (ii) The peak of light 
emission of the accretion disk lies in the blue region of the 
spectrum, so that the amplitude of the eclipse should increase
as the wavelength becomes shorter (``as the wave
length is increased'' in their text is mistaken).
However, they observed the opposite trend: 
the amplitude in the red region is much more deeper than
in the visual region.  As for the first conjecture (i), based on 
the accretion disk model together with the irradiation effects
of the disk and the companion, we have already constructed 
the orbital light curves that match well the observational data.
Second, we should be careful with the above statement (ii) because
the depth of the primary eclipse varies from night to night 
as seen in Figure 3 of \citet{kat04t} and its depth changes
as large as $\Delta R_c \sim 0.4$ between different periods of
the observations as shown in Figures 2 and 3.  Pavlenko et al.'s (2003)
$R$ and $I$ light curves are not simultaneous ones but taken
in different periods ($R$ is earlier than $I$).  The $V$ light
curve is reconstructed from Barsukova \& Goranskii's (2003) data.

They also rejected  possibility (3) based on the same statement as (ii)
discussed above.  However, their multi-color results may simply 
suggest the fact that the eclipse depth varies from period (night) 
to period (night).  
Therefore, we cannot conclude statement (ii)
only from the different depths at the different periods.

For the assumed WD luminosity of $L_{\rm WD}= 3,000~L_\sun$,
the distance modulus is obtained to be $(m-M)_R = 12.05$ 
and the corresponding distance is estimated to be $d \sim 1.4$ kpc 
(see Table \ref{system_parameters_V1494_Aql}).
However, this does not mean 
the real distance to V1494~Aql because we do not know the real
luminosity of $L_{\rm WD}$ at the time of the two observations.
For instance, if we adopt $L_{\rm WD}= 750~L_\sun$, 
the shape of the orbital light curve hardly changes but
the distance modulus becomes $(m-M)_R = 11.45$ and $d \sim 1$ kpc.



\acknowledgments
     We thank L. Kiss for providing us their V1494~Aql data.
\begin{deluxetable}{lll}
\tabletypesize{\scriptsize}
\tablecaption{Adopted system parameters of V1494 Aquilae
\label{system_parameters_V1494_Aql}}
\tablewidth{0pt}
\tablehead{
\colhead{parameter} & 
\colhead{symbol} & 
\colhead{present work}  
} 
\startdata
inclination angle & $i$ & $78.5\arcdeg$ \\
WD mass & $M_{\rm WD}$ & $1.0~M_\sun$ \\
WD luminosity & $L_{\rm WD}$ & $750 - 3,000~L_\sun$ \\
MS mass & $M_{\rm MS}$ & $0.3~M_\sun$ \\
mass accretion rate & $\dot M$ & $0.3 \times 10^{-8}~M_\sun$~yr$^{-1}$ \\
MS temperature & $T_{\rm MS,org}$ & 3,000~K \\
distance modulus & $(m-M)_R$ & $11.45 - 12.05$ \\
color excess & $E(B-V)$ & $0.60$\tablenotemark{a} \\
$R$-band absorption & $A_R$ & 1.39\tablenotemark{b} \\
distance & $d$ & $1.0 - 1.4$~kpc \\
\enddata
\tablenotetext{a}{taken from \citet{iij03}}
\tablenotetext{b}{from extinction law of \citet{rie85}}
\end{deluxetable}






\clearpage

\clearpage

\clearpage


\end{document}